\documentclass{mn2e}
\usepackage{psfig}
\usepackage[authoryear]{natbib}

\title{X-ray photometry}
\author[M.J. Page]{
M.J. Page\(^{1}\)
\\
\(^{1}\)Mullard Space Science Laboratory, University College London,
Holmbury St Mary, Dorking, Surrey RH5 6NT, UK.\\
}

\date{}

\begin{document}
\maketitle

\begin{abstract}
  I describe a method for synthesizing photometric passbands for use
  with current and future X-ray instruments. The method permits the
  standardisation of X-ray passbands and thus X-ray photometry between
  different instruments and missions. The method is illustrated by
  synthesizing a passband in the {\em XMM-Newton} EPIC pn which is
  similar to the {\em ROSAT} PSPC 0.5--2~keV band.
\end{abstract}

\section{Introduction}

Photometry, the measurement of source brightness in a given wavelength
range, is one of the most fundamental techniques in astronomy. It is
practised throughout the electromagnetic spectrum, and X-ray astronomy
is no exception. In optical astronomy, photometric passbands have been
standardised between observatories since the introduction of the
Johnson and Morgan system \citep{johnson53}; see \citet{bessel05} for a review. 
The benefits of a
well-defined set of photometric X-ray bands have been recognised
before, \citep[e.g. ][]{grimm09} but despite more than 50 years of
X-ray astronomy, there is as yet no system of photometric passbands
that applies to more than a single X-ray observatory. In this letter I
will describe the concept of synthesized X-ray passbands that can be
applied to existing and future X-ray astronomy instruments, and so permit the
establishment of standard, multi-mission X-ray photometric
passbands. As an example, I will show that the {\em ROSAT} PSPC 0.5-2
keV passband can be synthesized with the {\em XMM-Newton} EPIC pn
camera, so that X-ray photometry gathered with {\em ROSAT} and {\em
  XMM-Newton} can be compared directly. 

\section{Definitions}

I will adopt a definition of photometry as the measurement of source brightness
through a well defined passband relative to some standard
reference. The standard reference could be an astronomical object such
as the star Vega. Alternatively it could be a theoretical reference,
such as the constant flux density spectrum used as a reference in the
AB system \citep{oke65,oke74}, provided that the throughput of the
instrument performing the photometric measurements is sufficiently
well calibrated. Almost all measurements of source brightness in X-ray
astronomy may be considered to conform to such a definition: the
curves of effective area against energy commonly employed in X-ray
astronomy represent well defined passbands which have already been
calibrated in physical units. However, where X-ray photometry is disadvantaged
with respect to e.g. optical photometry, is that each instrument on
each X-ray observatory has a different effective area curve. 
To mitigate this problem, X-ray astronomers have developed tools such
as {\sc pimms}\footnote{https://heasarc.gsfc.nasa.gov/docs/software/tools/pimms.html}
 \citep{mukai93} which allow a measurement with one instrument to be translated into a
predicted count rate for a different instrument, given a model for the
X-ray spectrum of the source. However, if the spectrum of the source is not
known or is only poorly determined, then the predicted count rate will
not be accurate.

In a photon counting detector, as usually employed in X-ray astronomy instrumentation,
the count rate $C$ can be related to the spectrum of a source and
the effective area of the instrument as follows:
\begin{equation}
C = \int_{\nu_{min}}^{\nu_{max}} \frac{f_{\nu}(\nu)}{h\nu} A(\nu) d\nu
\end{equation}
where $\nu$ is frequency, $h$ is Planck's constant, 
$f_{\nu}(\nu)$ is the flux per unit
frequency interval of the source and $A(\nu)$ is the effective area of
the instrument. For an imaging system, it is assumed that $A(\nu)$ is
the overall effective area of the instrument and telescope combined. The
limits $\nu_{min}$ and $\nu_{max}$ are the minimum and maximum
frequencies at which $A(\nu)$ is significant.

If we define \[K={\rm max} \left(\frac{A(\nu)}{h\nu}\right)\] and 
\[R(\nu) = \frac{A(\nu)}{Kh\nu}\] so that $R(\nu)$ 
is the normalised respose curve of the system, then
\begin{equation}
\frac{C}{K} = \int_{\nu_{min}}^{\nu_{max}}f_{\nu}(\nu)R(\nu) d\nu
\label{eq:photometry}
\end{equation}

The fundamental quantity measured in astronomical photometry is the
quantity on the right hand side of Equation~\ref{eq:photometry}, the
integral over the passband of the flux of the source multiplied by the response
of the system. The quantity on the left hand side of
Equation~\ref{eq:photometry} is the measurement itself, the count rate
divided by the peak effective area of the system.
In optical astronomy the measurement is usually expressed as a magnitude $m$, 
\[m=-2.5\log_{10} \int_{\nu_{min}}^{\nu_{max}}f_{\nu}(\nu)R(\nu) d\nu + z\] 
where $z$ is the zeropoint, defined such that a reference
standard has zero magnitude \citep[e.g.][]{cousins76}. A Vega-based
magnitude system would not be appropriate in the X-ray regime, because
Vega has yet to be convincingly detected as an X-ray 
source \citep{ayres08,pease06},
but the AB magnitude system would be well suited to X-ray
astronomy. Nonetheless, the adoption of magnitudes as a convenient
unit is not necessary for the establishment of a system of photometric
passbands in X-ray astronomy. What {\em is} essential is the ability
to perform the measurement described by Equation~\ref{eq:photometry}
with different instruments on different observatories with similar
response functions, or passbands, $R(\nu)$.

\section{Synthesizing photometric passbands}

The synthesis of a passband is only possible for instruments which 
discriminate the energies of the incoming photons.
In X-ray astronomy, data are usually telemetered to Earth as `event
lists', in which the individual photons are listed with their
properties such as arrival time, position on the detector and energy
channel. The energy resolution of non-dispersive Charge-Coupled Device
(CCD) based X-ray instruments is limited, usually $10<R<50$ 
\citep[e.g. ][]{turner01,struder01}, so the
relationship between energy channel and (real) photon energy is
described by the redistribution matrix. The redistribution matrix
contains, for a sequence of discrete energy ranges, the probability of
a photon within that energy range being recorded in each of the energy
channels. I will refer to the product of the redistribution matrix and the
effective area as a function of energy (i.e. for each discrete energy
range, the channel probability distribution multiplied by the
effective area in that energy range) as the response
matrix\footnote{It should however be noted that, somewhat confusingly,
  it is quite common in X-ray astronomy to use the term `response
  matrix' interchangably with `redistribution matrix'.}. For each energy 
channel $i$, the response matrix contains the effective area for that 
channel as a function of energy, and therefore frequency $A_{i}(\nu)$.

X-ray astronomers choose passbands for their instruments by selecting
ranges of energy channels which correspond to their desired energy
ranges; the {\em Chandra} ACIS bands advocated by \citet{grimm09}, and
the bands chosen for the 2XMM catalogues \citep{watson09} are examples of this
approach, and I will refer to these bands as natural instrumental
passbands. For natural passbands, the count rate in the band
$C$ is related to the count rates in the individual energy channels
$C_{i}$ by
\[
C=\sum_{i_{min}}^{i_{max}}C_{i}
\]
where $i_{min}$ and $i_{max}$ are the minumum and maximum channels of the 
passband. The effective area of the band $A(\nu)$ is related to the effective areas $A_{i}(\nu)$ of the individual channels by
\[
A(\nu)=\sum_{i_{min}}^{i_{max}}A_{i}(\nu)
\]
While natural passbands can be chosen to cover similar energy ranges with different instruments on different missions, there is little scope in this approach to control (and thus standardize) the shapes $R(\nu)$ of the passbands. 

To synthesize a passband, it must be possible to control the shape as well as the energy range. This can be accomplished by assigning weights $w_{i}$ to the channels that form the band when calculating the count rate within the band. Thus for a synthesized band:
\[
C=\sum_{i_{min}}^{i_{max}}w_{i}C_{i}
\]
and
\[
A(\nu)=\sum_{i_{min}}^{i_{max}}w_{i}A_{i}(\nu)
\]
The weights can be chosen so that the synthesized band has a 
response $R(\nu)$ as close as possible to the response desired. The
weights $w_{i}$ therefore act as a synthetic filter, controlling the
shape of the passband. The capability to compose and apply the
synthetic filter after the data have been taken is a distinct
advantage of X-ray instrumentation compared to that employed for
photometry at longer wavelengths.

\section{Practical considerations}
\label{sec:considerations}

In order to synthesize a passband with an instrument, the instrument must be 
sensitive over a similar or larger frequency range than that covered by the passband. 
The degree to which the shape of a passband can be controlled by the
choice of the weights $w_{i}$ depends on the width of the channel
distribution at a given energy in the response matrix (i.e. the energy
resolution of the instrument) and to a lesser degree on the shape of
this distribution. 
The introduction of the weights
$w_{i}$ implies a reduction in the signal to noise of the count rate
with respect to that in a natural passband. In general, X-ray instruments
have improved in both spectral resolution and collecting area with
time, and this trend is expected to continue with the future 
European Space Agency mission
{\em Athena} \citep{nandra13}. 
Thus it is more practical to use present-day or future
instruments to synthesize the natural bands of earlier X-ray astronomy
instruments than the other way round.

In general, the more closely the response curves $R(\nu)$ of two
passbands match, the more similar the photometry will be between
them. Therefore the best photometric performance from a synthesized
band will be obtained by choosing the weights $w_{i}$ to match as
closely as possible the desired $R(\nu)$. On the other hand, the
smoother, and more uniform the $w_{i}$, the higher the signal to
noise will be. Therefore, in choosing the weights $w_{i}$, there
might be some trade-off between obtaining the desired shape $R(\nu)$
and maximising the signal to noise ratio, particularly if the best
match to $R(\nu)$ corresponds to large fluctuations in $w_{i}$
between adjacent channels.

Another consideration is that the count rate in a synthesized band
will no longer be Poisson distributed as it is in natural bands;
instead it will be distributed as the sum of the weighted Poisson
distributions, and error analysis is more involved. However, the
subtraction of background also causes the count rates of sources to
deviate from Poisson statistics, and this is routinely performed in
X-ray astronomy already, so the importance of this consideration
should not be exaggerated. Nonetheless, the reduction of signal to
noise and the increased complexity of error analysis are disadvantages
associated with synthesized passbands, to be traded off against the
advantages of standardising the passbands between missions. 

\section{An example synthesized passband}

As an example, I will now show that the 0.5-2 keV band of the {\em
  ROSAT} Position Sensitive Proportional Counter can be synthesized
with the {\em XMM-Newton} European Photon Imaging Camera (EPIC) pn
instrument \citep{struder01}. To date the only imaging all-sky survey
carried out in X-rays used the {\em ROSAT} PSPC, so this instrument
has provided photometric measurements of a large body of X-ray
sources \citep{voges99}. 
Many of these sources have been observed subsequently with
{\em XMM-Newton}. It would be useful to be able to compare directly the
photometric measurements from {\em ROSAT} and {\em XMM-Newton}, for
example to study the long term variability of X-ray sources. Following
the logic of Section~\ref{sec:considerations}, it is more practical to
synthesize the {\em ROSAT} PSPC passband with the XMM-Newton EPIC pn
than the other way around, because EPIC pn has better energy
resolution, and a larger effective area than the PSPC.

\begin{figure}
\begin{center}
\leavevmode
\hspace{-8mm}
\psfig{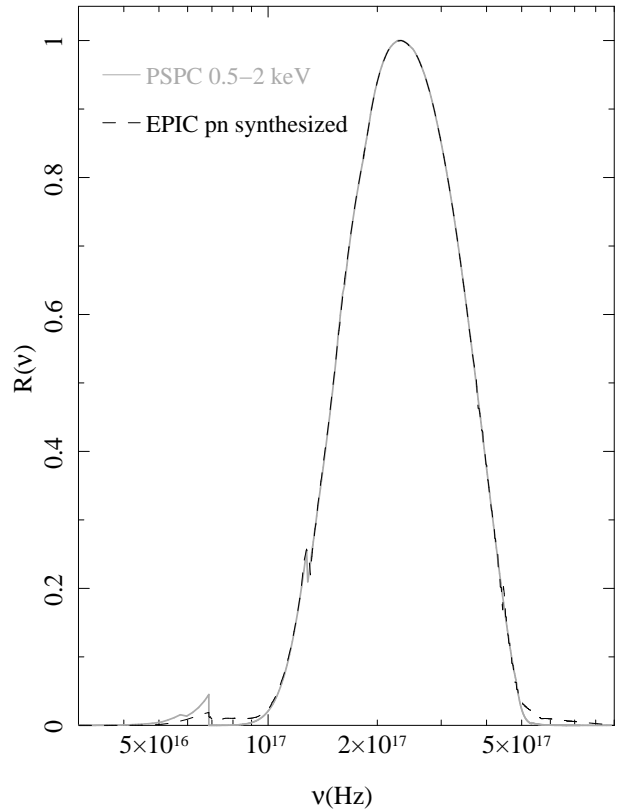}
\caption{Passband of the {\em ROSAT} PSPC 0.5 -- 2 keV band (solid
  curve) and an {\em XMM-Newton} EPIC-pn passband (dashed curve) which
  has been synthesized to have approximately the same shape.}
\label{fig:passbands}
\end{center}
\end{figure}

\begin{figure}
\begin{center}
\leavevmode
\hspace{-8mm}
\psfig{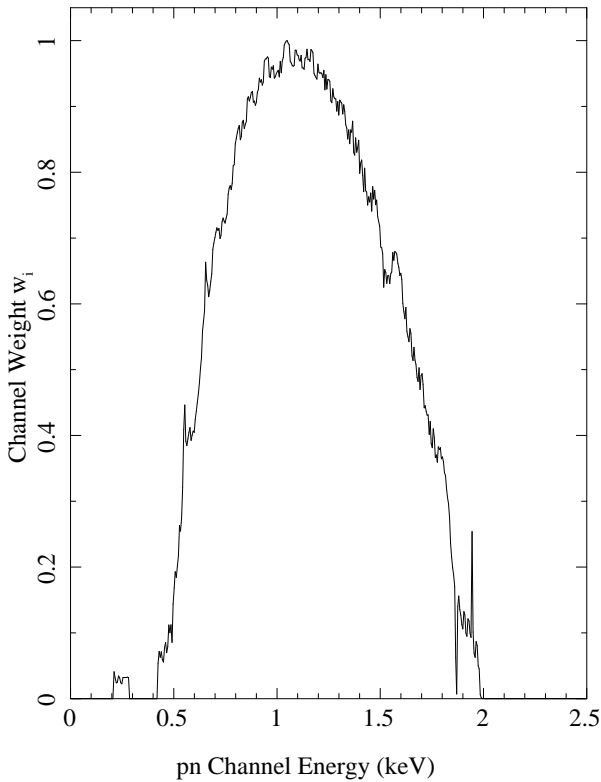}
\caption{Weighting function applied to the EPIC pn channels to produce the synthesized band shown in Fig.~\ref{fig:passbands}}
\label{fig:weights}
\end{center}
\end{figure}

The {\em ROSAT} PSPC 0.5-2 keV band is formed from PSPC energy
channels\footnote{These are usually called pulse-height invariant or
  PI channels, because they correspond to the pulse height of an event
  after it has been calibrated for temporal and spatial gain
  variations onto a standard energy scale} 52 -- 201.  The normalised
response $R(\nu)$ of this band is shown as the grey curve in
Fig.~\ref{fig:passbands}.  
The distribution of weights $w_{i}$ to be applied to the EPIC-pn 
channels was obtained by iteratively
adjusting the weights so as to minimize the sum of the squares of
the difference between the PSPC and synthesized response $R(\nu)$
at each point of the $R(\nu)$ curve
over the frequency range of the synthetic band, i.e. a least-squared
fit. No smoothing critera were applied during the fitting of the
weights.  Fig.~\ref{fig:weights} shows the channel weights $w_{i}$
corresponding to the synthesized passband. The distribution of
$w_{i}$ is quite smooth with fluctuations between adjacent channels
rarely exceeding 2 per cent, so smoothing the $w_{i}$ would have
little impact on the signal to noise ratio of photometry through the
synthesized band.  The synthesized $R(\nu)$ is shown as the black,
dashed curve in Fig.~\ref{fig:passbands}. The match between the {\em
ROSAT} PSPC and EPIC pn synthesized bands is good, with $R(\nu)$
differing by less than 0.01 over most of the band, and by no more than 
0.05 at any point. However, the
weak tail in the synthesized passband which stretches to higher
frequency is irreducible, because it corresponds to the low energy
tail\footnote{i.e. redistribution of photons of a given energy to
channels of lower energy} of the CCD response.

\begin{table*}
\caption{Simulated 0.5--2~keV photometry in the PSPC, EPIC-pn, and the synthesized PSPC band, for a number of spectral models. The second, third and fourth columns give the 0.5--2~keV flux that would be observed if the given spectrum is observed using the PSPC, EPIC-pn and Synthesized PSPC bandpass respectively, and the count rate is translated to a flux using a standard energy conversion factor. In all cases the spectral model was normalised such that the true 0.5--2~keV flux is 1.00$\times 10^{-14}$~erg~cm$^{-2}$~s$^{-1}$. The last two columns give the ratios of the flux that would be observed with EPIC-pn and the synthesized band respectively, to the flux that would be observed with the PSPC.}
\label{tab:simulations}
\begin{tabular}{lccccc}
Model&\multicolumn{3}{c}{--- Simulated Flux ($10^{-14}$~erg~cm$^{-2}$~s$^{-1}$) ---\ \ \ }&EPIC-pn /&Synthesized /\\ 
&\ \ \ \ PSPC\ \ \ \ &\ \ \ \ EPIC-pn\ \ \ \ &Synthesized&PSPC&PSPC\\
\hline
&&&&&\\
power law $\Gamma = 1$                           &0.96&0.96&0.96&1.00&1.00\\
power law $\Gamma = 2$                           &1.02&1.02&1.02&1.00&1.00\\
power law $\Gamma = 3$                           &1.04&1.04&1.03&1.00&0.99\\
power law $\Gamma = 2$, $N_{H}=10^{22}$~cm$^{-2}$&0.68&0.84&0.71&1.25&1.04\\
mekal $kT = 0.3$~keV                             &1.11&1.11&1.11&1.00&0.99\\
mekal $kT = 0.7$~keV                             &1.32&1.12&1.31&0.85&0.99\\
mekal $kT = 1.5$~keV                             &1.16&1.03&1.15&0.89&0.99\\
black body $kT = 0.3$~keV                         &1.09&1.01&1.08&0.93&0.99\\
\end{tabular}
\end{table*}

To investigate the performance of the synthesized band in reproducing
PSPC photometry, I have similated a number of spectral models and
folded them through the responses of the PSPC, EPIC-pn and the
synthesized 0.5--2 keV band using {\sc xspec11} \citep{arnaud96}. For
the normal EPIC-pn 0.5--2 keV response, I simply selected the pn
channels that have nominal energies within the range 0.5--2 keV. The
spectral models were chosen to span the range of spectral shapes
typically exhibited by sources in the 0.5--2 keV band, and comprise
power laws, with photon index $\Gamma = $1, 2 and 3, a power law with
$\Gamma = 2$, absorbed by cold gas with a column density of $N_{H} =
10^{22}$~cm$^{-2}$, optically thin thermal plasmas (mekal in {\sc
  xspec}) with $kT=$0.3, 0.7 and 1.5 keV, and a black body with
$kT=$0.3~keV. Each of the spectral models was normalised so as to have
the same true 0.5--2~keV flux of $10^{-14}$~erg~cm$^{-2}$~s$^{-1}$,
and the count-rates were simulated with zero statistical error, so
that any differences in the photometry from the 3 passbands are
systematic differences related to the shapes of of the passbands.

In X-ray astronomy, photometry is normally derived from
imaging surveys by measuring the count rate in a given band, and
dividing this by an ``energy conversion factor'' (ECF) which is
specific to the instrument and band to obtain a flux in physical
units. We have followed this process in simulating the photometry that
would be obtained through the different passbands. The ECF itself is
typically derived by folding a simulated source of known flux and
spectral shape through the response of the instrument following
the same procedure as we are employing to obtain the count
rates for our simulated sources. To generate ECFs for the three
passbands, we adopt the spectral model used for the 3XMM catalogue, a
power law of photon index $\Gamma = 1.7$ absorbed by cold gas with a
column density $N_{H}=3\times 10^{20}$~cm$^{-2}$ \citep{rosen15}. 
The ECFs for the PSPC, EPIC-pn and the synthesized band are 
$7.65\times 10^{10}$, $6.67\times 10^{11}$ and 
$4.75\times 10^{11}$ ergs$^{-1}$~cm$^{2}$ respectively.

The results of the simulated photometry are given in
Table~\ref{tab:simulations}. It can be seen that the flux derived from
the PSPC differs significantly for several spectral models from the
flux derived from photometry in the natural EPIC-pn 0.5--2~keV
band. In the worst case of the heavily absorbed spectrum, the flux
differs by 25 per cent. In contrast, the flux derived from the
synthesized band differs from the PSPC flux by less than 1 per cent
for all spectral models except for the strongly absorbed spectrum. In
that case the difference is only 4 per cent, and is due to the weak
tail towards higher energy in the synthesized response, where the
absorbed spectrum is rising rapidly.

\section{Conclusions}
\label{sec:conclusions}

I have outlined a method for synthesizing photometric passbands in
X-ray astronomy. It offers the potential to standardize photometry
from different X-ray astronomy instruments and missions. I have
limited this letter to a conceptual description of the approach, and
provided an example to illustrate how it might be used.  Widespread,
practical application of the method will require considerably more
work. Software tools will be required to optimize the shapes of the
synthetic filters $w_{i}$ used to shape the passbands, and software
tools will be needed to compute photometry for real astronomical
sources in the synthetic passbands. Not least, for this method to
facilitate the large-scale provision of photometry in standardized X-ray
passbands, there will need to be concordance within the X-ray
astronomy community, at least amongst the curators of archives and the
constructors of astronomical catalogues, as to what to choose as the
minimum set of standard passbands.

\section{Acknowledgments}

I thank Mark Cropper and Francisco Carrera for useful discussion.

\end{document}